# MERGING OF THE GRAINS DURING WIRE DRAWING

L. Metlov[1,2], A. Zavdoveev[1,3*], E. Pashinska[1,2]

[1]Donetsk Institute for Physics and Engineering named after A.A. Galkin of the NAS of Ukraine, Prospect Nauky, 46 , Kyiv, Ukraine, 03028
[2]Donetsk national university, Universitetska, 24, Donetsk, 83001, Ukraine
[3]Paton Electric Welding Institute of NAS of Ukraine, Bozhenko, 11, Kiev, 03680 Ukraine
*avzavdoveev@gmail.com

Abstract

It has been first proved the effect of grains merging during drawing deformation. This was done with example of producing a steel wire from rod manufactured by rolling with shear technology and was shown not only grain refinement but its merging as well. The result obtained in current work has fundamental importance; it reveals new mechanism of the "recrystallization" which takes place without diffusion actions owing to the mechanical impact.

## 1. INTRODUCTION

Severe plastic deformation (SPD) initiates the generation of structural defects such as dislocations and grain boundaries in the first place. Increasing the number of grain boundaries is automatically lead to grain refinement of last. At the same time, along with the processes of grain refinement the opposing processes of their consolidation can occur. It is believed that such processes can result in the classical recrystallization involving atom- wise diffusion. However atom- wise diffusion is a slow process and its role can be significant only in the processes of aging at high temperatures. In the rapid processes, such as SPD, at low temperatures can be expected from the contribution from other mechanisms of recrystallization.

In ref.[1] recrystallization processes are divided into continuous and descrete recrystallization. The first of these is performed through nonthreshold mechanism and include such steps as the nucleation and growth of grains. The second of these is performed through threshold mechanism as a result of the continuous restructuring of the metals structure. Restructuring accompanied with texture rotation of the grains so that their crystallographic planes become collinear, thereby eliminating the need for additional grain boundaries, separating the grains with different crystallographic directions of the plane. This mechanism has been confirmed experimentally in [2].

Previously it have been identified and studied the phenomenon of diffusionless merging of grains by molecular dynamics simulations of nanowire drawing process, that actually is diffusionless recrystallization [3, 4], which has the same nature as that in [1]. In the computer experiment was simulated stretching of nanowire consisting of two alternating nanograins with mutually orthogonal oriented axes of symmetry. In the drawing process the grain which axes direction are not optimally oriented with respect to the tensile force becomes unstable and continuously rebuilt one's structure. As a result, all axis of symmetry becomes collinear and grains are merging (enlarging) without diffusion processes. Direct experiments on the stretching wire, i.e. confirming the feasibility of such a mechanism until now the authors are not known, and will be presented here, perhaps for the first time.

In the literature [5- 8] extensively discussed the use of special methods of severe plastic deformation (SPD), which allow to receive structural features that provide high ductility and conductivity of the material without loss of the strength. One example of such treatment is a rolling with shear (RS), which is designed with the modern concepts of the mechanism of structure formation under SPD [9-11]. The concepts of thermodynamic role of shear stress for the structural adjustment of solids and for the generation of microcracks were made even earlier by one of the authors [12 - 14]. At the initial stage, these ideas were also the basis for the processing method of rolling with shear.

2. METHODOLOGICAL DETAILS.

The program material consisted of commercial grade low-carbon steel whose nominal chemical composition is shown in Table 1.

Table 1. Chemical composition of the low-carbon steel Grade 08G2S

| C | Mn | Si | S | P | Cr | Ni | Cu | N2 |
|---|---|---|---|---|---|---|---|---|
| 0,08 | 1,87 | 0,82 | 0,020 | 0,022 | 0,02 | 0,02 | 0,02 | 0,007 |

The first processing step was carried out by rolling and the second one cold drawing.

Step 1. Prior to rolling process the billets was heated to a temperature of 1200°C. Thereafter, the rolling was conducted through two technologies: the standard (ST -technology) and proposed (RS-technology) in [9-11]. During rolling with shear technology the number of passes and cooling conditions was identical to the standard technology.

The scheme of hot rolling with shear is realized. It is characterized by significant shear deformations in the roll groove during formation of the metal. Plastic deformation of metal is accomplished at least twice at the temperature below the lowest critical point of phase transformations. The partial reduction is not less than 0,10 and take place in the pairs of roll's grooves of simple form where the first is a seam pass and the second one is edging. The succeeding cooling are fulfilled in cooling medium at the rate not less than 1,5 ° C/min down to the temperature of the end of the structure transformations. A peculiarity of the scheme is that rolling in edging passes is made with grooves displaced along the axis of rolls at the distance of 0,05 ... 0,20 of the groove width [10].

The deformation accumulated at the rolling with shear can be decomposed into too components: the deformation of rolling and deformation of shear respectively. To estimate the rolling deformation, we used the formula e=ln(S/S$_0$). In our case the value of *e=0.36*. Then we calculated the shear deformation, using the formula for deformation under ECAP, exactly $e = 2\frac{ctg\Theta}{\sqrt{3}}$ [7].

As the deformation zone is inhomogeneous, we should integrate this expression over the whole θ. Obtained average value of deformation, accumulated in the shifting pass *e=0.14*. The total accumulated strain is *e = 0.36+0.14 = 0.5*, that is 40% higher the deformation without shear. The estimated values are agreed satisfactory with measurements of the power consumption of shear rolls engines [5]. It is well known that the power consumption is proportional to the pressure on the operating tool and accumulated strain. In [5], the registered difference of power consumption is about 30% that is 10% less than changes of the deformation according to our experimental estimates. The divergence is quite acceptable with respect to the formula adapted to ECAP, which was used for strain evaluation.

Step2. The rod of 6 mm in diameter, obtained by ST-technology, was drawn at ambient temperature to 2.3 mm (cold drawing). Upon reaching 3.4 mm the drawing process became impossible due to the loss of ductility of the wire and, as a consequence, further numerous breakage. Therefore, to recover ductility of the wire the intermediate softening annealing was done in a tube furnace at a temperature of 600°C one hour with air cooling. The rod of 9.15 mm in diameter, obtained by RS-technology, was drawn at ambient temperature to 1.55 mm without intermediate softening annealing.

Microstructural observations were performed by optical microscopy and electron backscatter diffraction (EBSD). For optical microscopy, the microstructural specimens were mechanically polished in conventional fashion and finally chemically etched by using a 4% Nital solution. A final surface finish for EBSD was obtained by electro-polishing in a solution of 65% orthophosphoric acid + 15% sulphuric acid + 6% chromic anhydride + 14% water. The important electro-polishing parameters included: temperature 70-90°C, anodic current density 1 A/cm$^2$, voltage 23 V, and exposure 19 s.

EBSD analysis was conducted using a JSM-6490LV scanning-electron microscope equipped with HKL EBSD software. Depending on particular material condition, orientation mapping was performed using scan step size of 0.3 or 0.5 μm. To improve the reliability of the EBSD data, EBSD maps were "cleaned" using standard clean-up option of the HKL software. In addition, to eliminate spurious boundaries caused by orientation noise, a lower-limit boundary-misorientation cut-off of 2° was used. A 10° criterion was used to differentiate low-angle boundaries (LABs) and high-angle boundaries (HABs)

## 3. RESULTS AND DISCUSSION

Studies of the initial microstructure of steel rod 08G2S showed that the structure is a homogeneous mixture of ferrite and pearlite. Microstructure of the rod of 6 mm diam. obtained by ST - technology represents equiaxed ferrite and pearlite grains, the average grain size of 9.5 microns (cross section). In the longitudinal direction of the samples the elongation of the grains are observed, average grain size of 14 microns [see ref. 8]. **With** previous application of RS-technology followed by application of cold drawing is obtained completely different structure than in the case of ST-technology. For example, a combination of rolling and cold drawing gives a fine structure with a large number of fine grains, while the use of RS-technology and subsequent cold drawing (Fig. 1.b) leads to the appearance of large grains together with small ones. The latter is due to the peculiarities of the formation of structure in the rolling with shear process. This structure inherited during subsequent deformation processing, and is manifested in the cyclic changes of grain sizes with strain accumulation, due to the intensified movement defects (IMD) [15]. The processes of dynamic recrystallization, and as a result, grain coarsening develop more intensely.

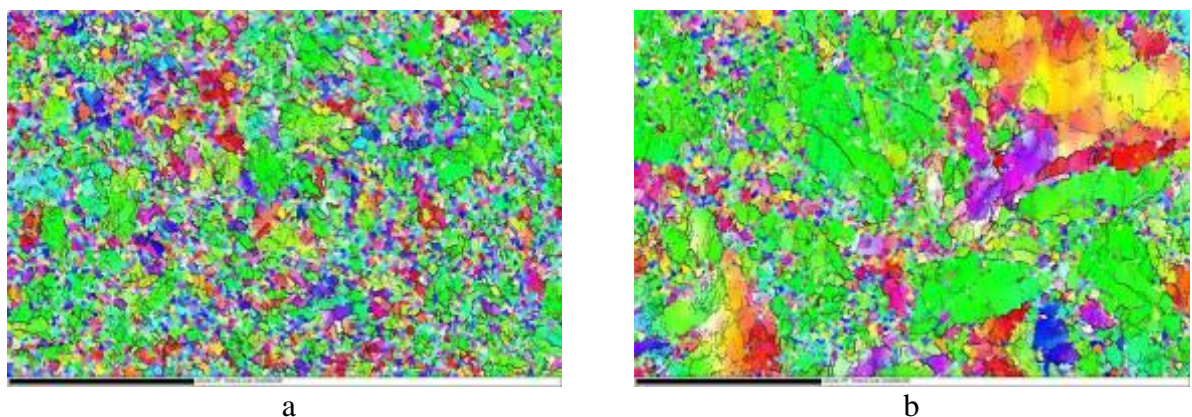

Fig.1. EBSD maps of low carbon steel after rolling with subsequent drawing. a – ST-technology + cold drawing, b – RS-technology + cold drawing. Scale bar – 20 μm

The prevalence of orientation (101) is strongly expressed in the samples subjected to hot rolling followed by drawing (Fig. 1 and, b). This fact is easy to explain: as in bcc metals the main slip plane is (101), the observed increase in the area of green areas on the orientation map has deformation nature.

From Fig. 2. a, t is seen that with increasing degree of deformation for ST - technology decreases the average grain size in cross section and increases the grain size in a longitudinal section, i.e. in the direction of the axis of deformation. There has been some increase in <$d_{long}$> at the last stage of deformation ⌀ 3,5 → 3.4 mm respectively, due to the annealing performed at this stage of drawing.

Actual growth of grains at the initial stage of drawing is reflected in Fig. 2, b. Thus, under intense deformation, which is a drawing, not only grain refinement, as usually considered, but even on the contrary, grain enlarging is observed. They decrease in transverse dimension that creates the illusion of crushing, but elongate in the direction of the tensile force, which was paid no attention. The total effect is multiplexed absolute increase in the volume of grain.

To more clearly refine this effect is of interest to see how changing the actual size of the grain compared to their theoretical change in accordance with the hypothesis Polanyi-Taylor, that is, in proportion to the total deformation of the wire. To do so, we have introduced the concept of the relative size and the relative length of the grains that defined as follows: $Dr = \dfrac{Dav}{Dpt}$, where $Dav$ - The average grain size in the cross section of the wire, $Dpt$ - the estimated size of the grain, according to the principle of Polanyi - Taylor ( $Dpt = D_0 \cdot \exp^{(-\frac{e}{2})}$ ), were $D_0$ - the initial grain size, $e$ - deformation of the drawing.

$Lr = \dfrac{Lav}{Lpt}$, where $Lav$ - the average grain size in longitudinal section of the wire, $Lpt$ - the estimated length of grain, according to the principle of Polanyi - Taylor ( $Lpt = L_0 \cdot (\exp^{(e)} + 1)$ ), where $L_0$ - initial grain size, $e$ - deformation of the drawing.

From Fig. 1 c,d it is seen that higher relative grain diameter higher and grows at an initial stage of drawing compared with its theoretical value according to the hypothesis Polanyi - Taylor, while the relative length there of, on the contrary, decreases. This suggests that the increase of the relative grain diameter associated with the phenomenon of texture merger, so reducing the diameter of the grain is much slower theoretical. At the same time, the increase in the average grain length less than the theoretical. This indicates that if the grain fusion and axially stretching occurs, it is largely compensated by an oppositely directed grains fracture process that prevails here.

4. CONCLUSIONS REMARK.

Thus, in the RS- technology the fact of the average amount of grain growth at the initial stage of rolling is recorded. This effect is conditioned by intensive grows of transverse dimension of grains (EBSD maps Figure 1) and can be explained by the diffusionless merging of the grains due to texture turning of grains. The obtained result has fundamental importance, it reveals a new mechanism of "recrystallization", which is carried out without the participation of diffusion processes because only the mechanical components of the action of a force field.

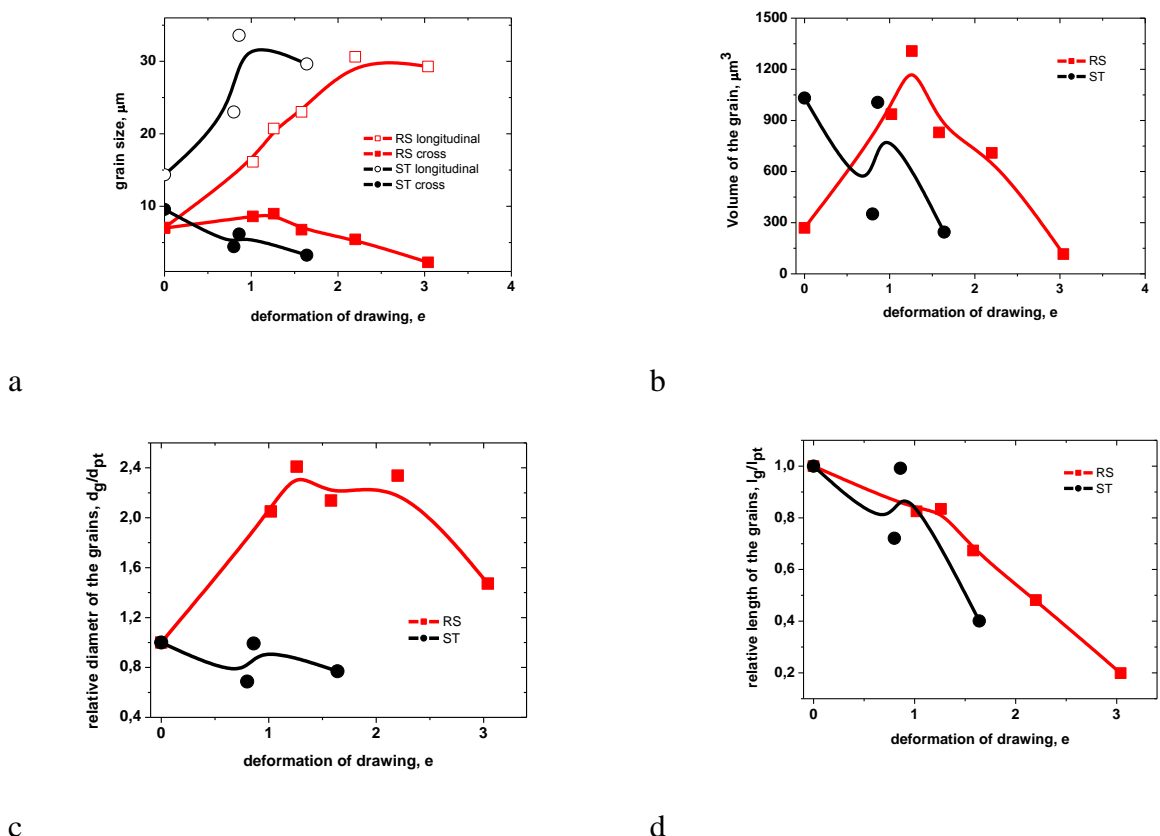

Fig. 2. Quantitative analysis of microstructural data. a – grain size distribution, b – volume of the grains, c – relative diameter of the grains, d – relative length of the grains. RS – Rolling with shear technology with subsequent cold drawing, ST – Standard technology with subsequent cold drawing.